\documentclass{webofc}

\usepackage[varg]{txfonts}   
\usepackage{hyperref}
\usepackage{url}
\usepackage{lineno, blindtext, indentfirst}
\usepackage{graphicx, float}
\hypersetup{colorlinks=true,citecolor=blue,urlcolor=blue,linkcolor=blue}

\newcommand{ \pXi }{$p$$-$$\Xi^{-}$}
\newcommand{ \apaXi }{$\bar{p}$$-$$\bar{\Xi}^{+}$}
\begin{document}
    
%
\title{Measurement of \pXi{} (\apaXi{}) Correlation Function in Isobar and Au+Au Collisions at $\sqrt{s_{\mathrm{NN}}}$ = 200 GeV with the STAR Detector}
%
%

\author{\firstname{Boyang} \lastname{Fu for the STAR Collaboration}\inst{1}\fnsep\thanks{\email{boyangfu@mails.ccnu.edu.cn}} 
}

\institute{Institute of Particle Physics and Key Laboratory of Quark \& Lepton Physics (MOE), \\Central China Normal University, Wuhan, 430079, China. 
\
          }

\abstract{ Understanding the strong interactions between baryons, especially hyperon-nucleon ($Y$-$N$) interactions, is crucial for comprehending the equation-of-state (EoS) of the nuclear matter and inner structure of neutron star. In these proceedings, we present the measurements of \pXi{} (\apaXi{}) correlation functions with high statistics in Isobar (Ru+Ru, Zr+Zr) and Au+Au collisions at $\sqrt{s_{\mathrm{NN}}}$ = 200 GeV by the STAR experiment. With the Lednický-Lyuboshitz approach, the source size and strong interaction parameters of \pXi{} (\apaXi{}) pairs are extracted. 
}
\maketitle
\section{Introduction}
\label{intro}
\setlength{\parindent}{0em}
Baryon interactions are of fundamental interest in nuclear physics and astrophysics. Further, the EoS of the nuclear matter at high baryon density that governs the properties inside neutron stars depends on baryon interactions [1]. Among all the baryon interactions, interactions between $Y$-$N$ and hyperon-hyperon ($Y$-$Y$) are of utmost importance to solve the hyperon puzzle [2]. \pXi{} (\apaXi{}) interactions with strangeness $S = -2$ sector have long been of interest in theory and experiments as they are proposed to be a potential decay channel for the H-dibaryon (consisting of six quarks: $uuddss$) particle [3]. 

\label{sec-2}
In heavy-ion collisions, two-particle femtoscopy is a powerful and unique method for extracting information about the spatio-temporal properties of the source, characterising the final state interactions (FSI), and searching for the possible bound states. In these proceedings, we present new results of \pXi{} (\apaXi{}) correlation using high statistics data from Isobar and Au+Au collisions at $\sqrt{s_{\mathrm{NN}}}$ = 200 GeV by the STAR experiment. In the following, we will only use \pXi{} to denote the \pXi{} and \apaXi{} pairs.  

\section{Correlation Function}
The observable of interest in femtoscopy is the two-particle correlation function. The correlation function can be expressed theoretically as $C(k^{*})=\int d^{3}r^{*}S(r^{*})\lvert \Psi(r^{*},k^{*})\rvert^{2}$, where $k^{*}$ and $r^{*}$ are the relative momentum and relative distance of the pair of interest in the Pair rest Frame (PRF) reference. $S(r^{*})$ is the source function and $\Psi(r^{*},k^{*})$ represents the wave function of the pair of interest. Experimentally, this correlation function is computed as $C(k^{*})=\mathcal{N}[A(k^{*})/B(k^{*})]$, where $A(k^{*})$ are the correlated pairs in same-event, and $B(k^{*})$ is the uncorrelated background pairs obtained from mixed-events. The normalization parameter $\mathcal{N}$ is chosen such that the mean value of the correlation function equals unity for $k^{*}$ [200, 400] MeV/$c$.

\label{sec-2}
The correlation function is parameterized using the Lednický-Lyuboshitz (LL) model [4], which considers a static spherical Gaussian source under a smoothness approximation [5], convoluted with an S-wave function. In LL model, the complex scattering amplitude $f^{S}(k^*)$ with Coulomb interaction is evaluated via the effective range approximation,
\begin{equation}
   f^{S}(k^{*})=\left[\frac{1}{f_{0}^{S}}+\frac{1}{2}d_{0}^{S}k^{*2}-\frac{2}{a_{c}}h(\eta)-ik^{*}A_{c}(\eta)\right]^{-1}
\end{equation}
where $f_{0}^{S}$ is scattering length, $d_{0}^{S}$ is effective range of the interaction, $a_{c}$ is the Bohr radius, $h(\eta)$ is complex function, $\eta$ and $A_{c}$ are Coulomb related factor, and $S$ denotes the total spin of the pair. A spin-averaged fit is considered in this work.

\section{Results}
\label{results}
\begin{figure}[h]
\vspace{-0.5cm}
\centering
\includegraphics[width=0.87\textwidth]{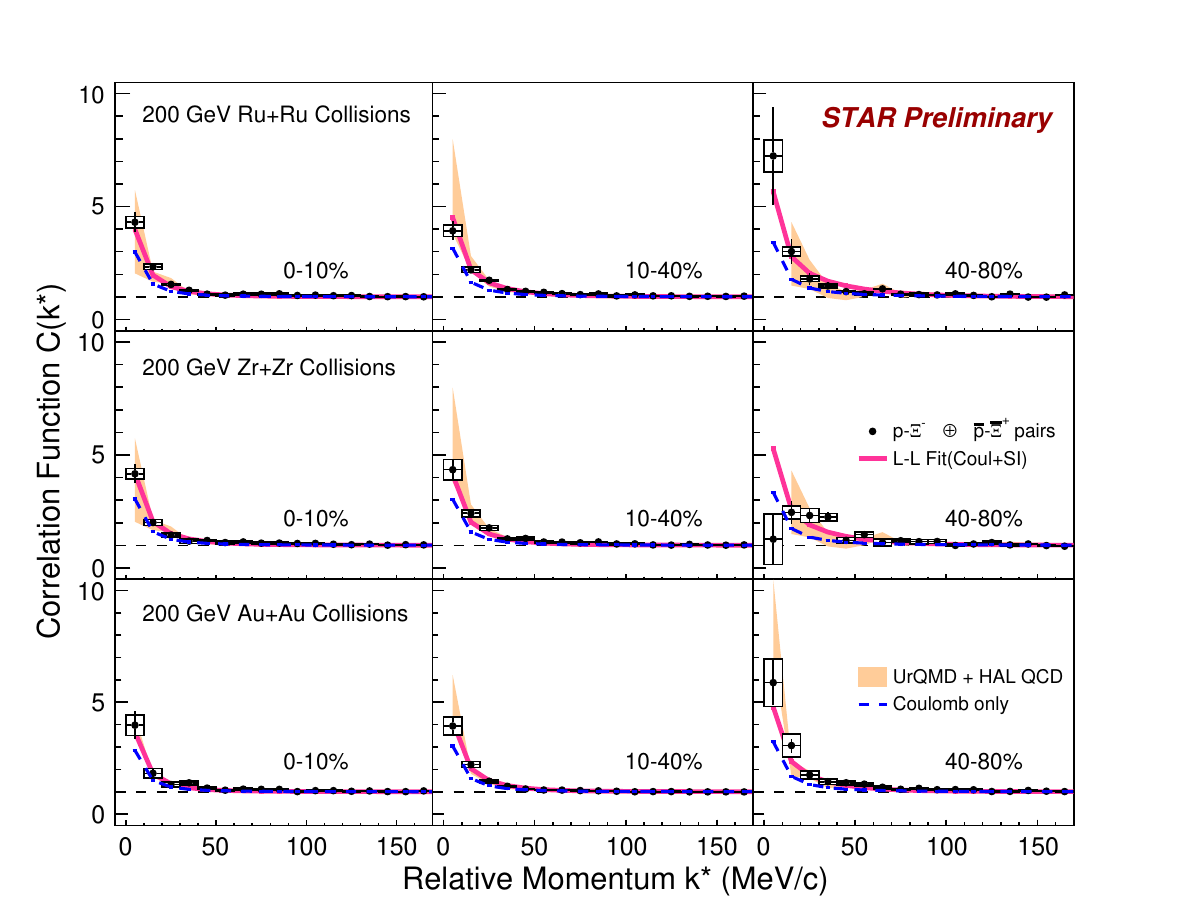}
\caption{Measured \pXi{} correlation functions in 0-10\% (left), 10-40\% (middle) and 40-80\% (right) centrality in Ru+Ru (top), Zr+Zr (middle) and Au+Au (bottom) collisions at $\sqrt{s_{\mathrm{NN}}}$ = 200 GeV. The black vertical bars and boxes represent the statistical and systematic uncertainties, respectively. In each colliding system, the measured correlation function of three centralities are fitted simultaneously with Lednický-Lyuboshitz function, shown as magenta lines, including effects of Coulomb and strong interactions. Correlation functions with pure Coulomb are shown as blue dashed lines. The orange bands represent the results from UrQMD + HALQCD simulation [6-7].}
\vspace{-0.4cm}
\label{fig:CF}       
\end{figure}

\setlength{\parindent}{0em}
Figure~\ref{fig:CF} shows the measured \pXi{} correlation function as a function of $k^*$ in three centrality bins in Isobar and Au+Au collisions at $\sqrt{s_{\mathrm{NN}}}$ = 200 GeV. In this analysis, the residual correlation from $p$$-$$\Xi(1530)$ has been removed and we neglect other pairs contribution. The measured correlation functions show clear enhancement at low $k^*$ in all centrality classes and become more pronounced in peripheral (40\%-80\%) collisions. The correlation functions are simultaneously fitted with LL model with five free parameters (Source size ($R_{G}$) for three centrality bins, a common $f_0$ and $d_0$) in each collision system using Bayesian method [8]. It is clear that only Coulomb interaction cannot account for the full measured correlation functions. Results of the UrQMD + HALQCD calculations [6-7] are shown as orange band in the Fig.~\ref{fig:CF}. The transverse momentum ($p_{T}$) spectra of single particles from the UrQMD model is employed to generate phase space, the interaction potential is provided by Lattice QCD [6-7]. The simulated model results describe the data reasonably well in all centrality classes. 
 
\begin{figure}[htbp]
\vspace{-0.1cm}
\centering
\includegraphics[width=0.67\textwidth,clip]{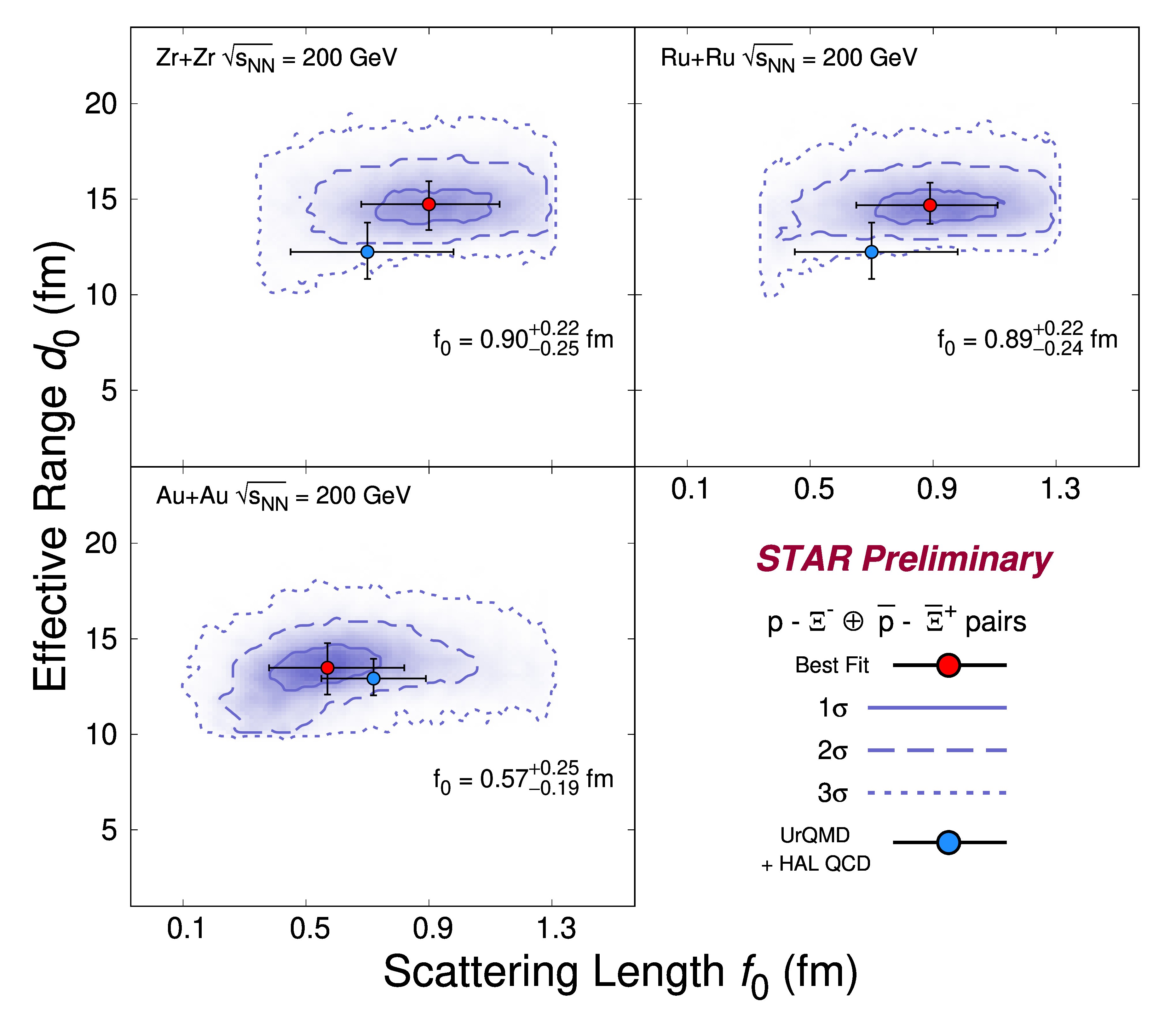}
\caption{
The \pXi{} spin-averaged scattering parameters ($f_0$) and effective range ($d_0$) obtained from simultaneous fit in Isobar and Au+Au collisions by Bayesian method. The best fit results are shown in red solid points. The purple bands and dashed lines represent the 1$\sigma$ to 3$\sigma$ standard deviations of the fit. The blue solid points represent the $f_0$ and $d_0$ extracted from UrQMD + HALQCD simulation with the same method.}
\vspace{-0.5cm}
\label{fig:contour}     
\end{figure}

\begin{figure}[htp]
\centering
\vspace{0.0cm}
\hspace{-0.5cm}
\includegraphics[width=0.5\textwidth,clip]{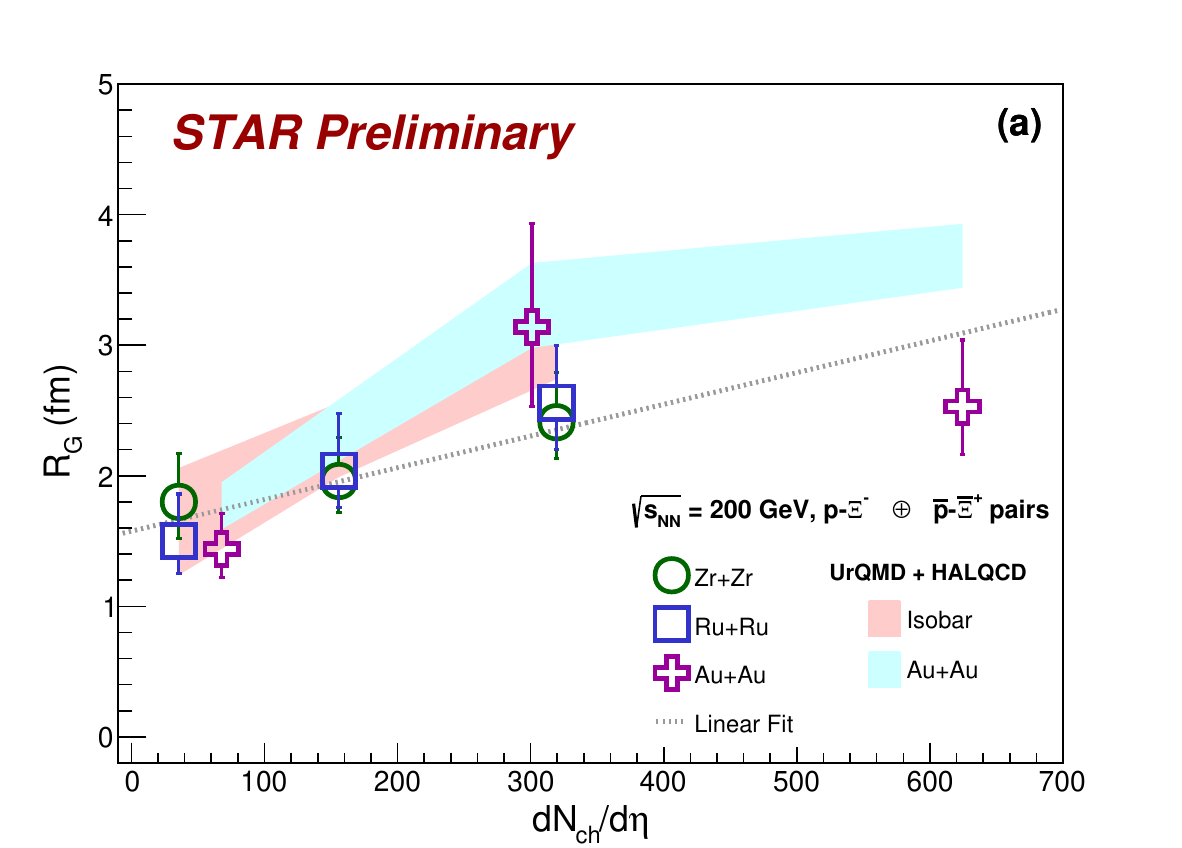}
\includegraphics[width=0.5\textwidth,clip]{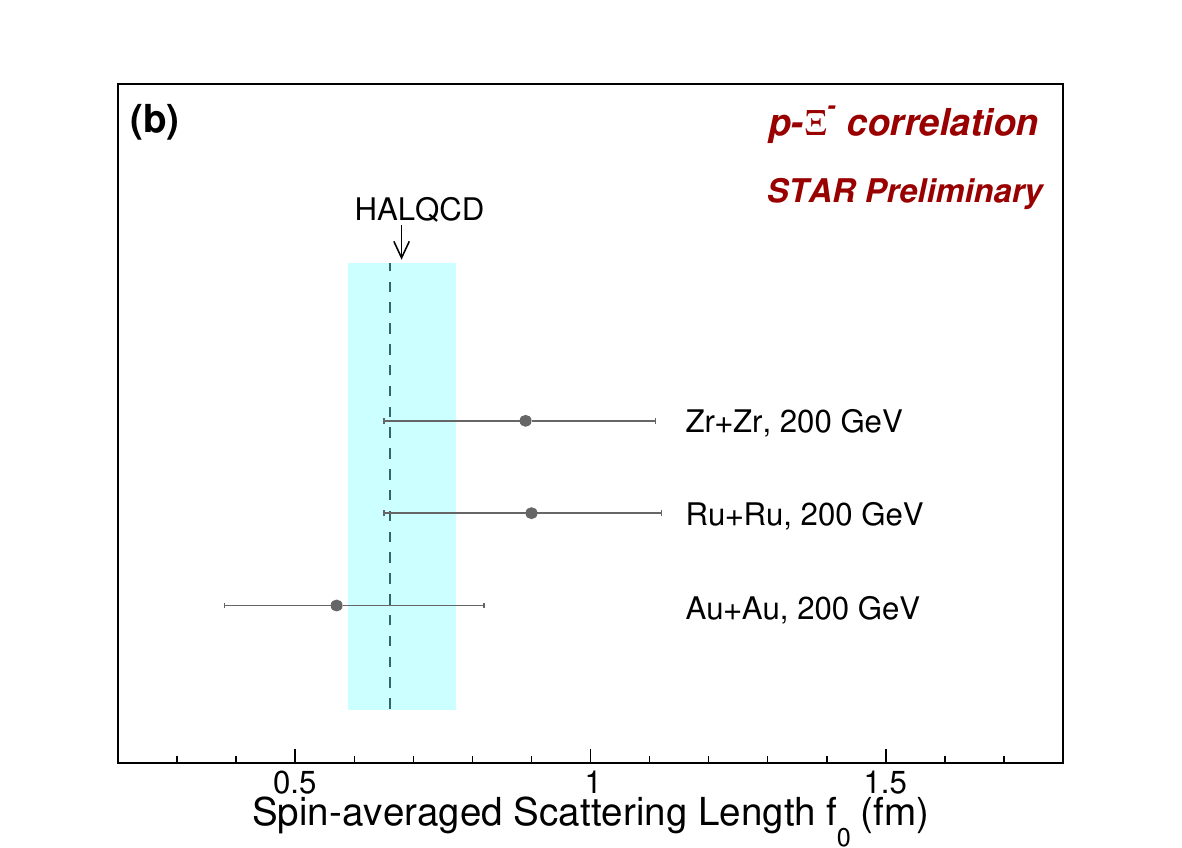}
\caption{(a) \pXi{} source size $R_G$ extracted from LL function as a function of $dN_{ch}/d\eta$ in Isobar and Au+Au collisions. The gray dotted line shows the linear fit to all data points. The pink and blue bands represent the $R_G$ extracted from UrQMD + HALQCD simulation. (b) Comparison of \pXi{} scattering length $f_0$ between experimental data and HALQCD theory predictions. The black points represent extracted $f_0$ from Ru+Ru, Zr+Zr and Au+Au collisions at $\sqrt{s_{\mathrm{NN}}}$ = 200 GeV. The blue band represent theory prediction from HALQCD Collaboration, the width of band represents the statistical uncertainties [6].}
\vspace{-0.5cm}
\label{fig:source}     
\end{figure}

\setlength{\parindent}{2em}
Figure~\ref{fig:contour} shows the probability density distribution of the strong interaction parameters $f_0$ and $d_0$ for \pXi{} pairs for each collision systems obtained from the LL fit. It is found that the $f_0$ and $d_0$ are consistent with Isobar and Au+Au collisions within 1$\sigma$. Using same fitting procedure, the results from UrQMD + HALQCD simulation are consistent with data within 1$\sigma$. The $f_0$ is determined in the range [0.57, 0.90] fm, which is relatively small, indicating that the attractive interaction between \pXi{} pairs is shallow. Figure~\ref{fig:source} panel (b) shows the comparison with HALQCD theory, it is observed that the HALQCD prediction is within good agreement with experimental data. This is the first time, the strong interaction parameters for \pXi{} pairs are extracted.

The obtained source size $R_G$ for \pXi{} pairs in different centrality classes from the LL fit are presented in Fig.~\ref{fig:source} panel (a). It is observed that the $R_G$ exhibit a monotonic decrease from central to peripheral collisions. Among different collision systems, the extracted $R_G$ tends to follow a linear distribution as a function of $dN_{ch}/d\eta$ (charged particles rapidity density). The same fitting were performed to UrQMD simulations and the resulting radius from model are close to data.

\section{Summary}
\label{summary}
\setlength{\parindent}{0em}
In these proceedings, we present the systematic measurements of \pXi{} correlation functions in Isobar and Au+Au collisions at $\sqrt{s_{\mathrm{NN}}}$ = 200 GeV. For the first time, the source size and strong interaction parameters are extracted for \pXi{} pairs in each collision systems with Lednický-Lyuboshitz approach. It is observed that, from central to peripheral collisions, the $R_G$ decreases. Calculations from the transport model UrQMD coupled with Lattice QCD potential provide a good description of the data. The extracted strong interaction parameters ($f_0$, $d_0$) are consistent between Isobar and Au+Au collisions. The $f_0$ has a rather small value indicating there is a shallow (Compared to typical $N$-$N$ interaction) attractive interaction in \pXi{} pairs. Our measurements provide strong constraints towards the understanding of $Y$-$N$ interactions.

\section{Acknowledgements}
This work was supported in part by National Key Research and Development Program of China (No.2022YFA1604900, No.2020YFE0202002), National Natural Science Foundation of China (No.12122505) and the Fundamental Research Funds for the Central Universities (CCNU220N003).

%


\begin{thebibliography}{}
%
%

\bibitem{RefJ}
Lisa et al, Ann.Rev.Nucl.Part.Sci. \textbf{55}, 357--402 (2005)

\bibitem{RefJ}
Lattimer et al, Science \textbf{304}, 536 (2004)

\bibitem{RefJ}
Jaffe, Phys.Rev.Lett. \textbf{38}, 195 (1977)

\bibitem{RefJ}
Lednicky et al, Sov.J.Nucl.Phys. \textbf{35}, 770 (1982)


\bibitem{RefJ}
Pratt et al, Phys.Rev.C \textbf{56}, 1095 (1997)


\bibitem{RefJ}
Kamiya et al, Phys.Rev.C \textbf{105}, 014915 (2022)

\bibitem{RefJ}
Sasaki et al, Nucl.Phys.A \textbf{998} 121737 (2020) 

\bibitem{RefJ}
M\"antysaari et al, Phys.Lett.B \textbf{833}, 137348 (2022)


\end{thebibliography}
\end{document}